\documentclass[journal]{IEEEtran}
\usepackage{amssymb}
\usepackage{cite}
\usepackage[cmex10]{amsmath}
\usepackage{graphicx}
\usepackage{amsmath}
\usepackage{epstopdf}
\usepackage{bm}
\usepackage{color}

\usepackage{algorithm}
\usepackage{algorithmic}
\ifCLASSOPTIONcompsoc
\else
  \usepackage[caption=false,font=footnotesize]{subfig}
\fi

\usepackage{setspace}
\usepackage{multirow}

\usepackage{amsfonts}

\begin{document}
% paper title
% can use linebreaks \\ within to get better formatting as desired
% Do not put math or special symbols in the title.
\title{Achievable Rate Region of MISO Interference Channel Aided by Intelligent Reflecting Surface}
%\author{\IEEEauthorblockN{Wei Huang\IEEEauthorrefmark{1}, Yong Zeng\IEEEauthorrefmark{2},  and Yongming Huang\IEEEauthorrefmark{2} }
%\IEEEauthorblockA{\IEEEauthorrefmark{1}School of Computer Science and Information Engineering, Hefei University of Technology, Hefei, China\\
%\IEEEauthorrefmark{2}National Mobile Communications Research Laboratory, Southeast University, Nanjing, China. }
%Email: huangwei@hfut.edu.cn, \{yong$\_$zeng, huangym\}@seu.edu.cn
%lu.zhaohua@zte.com.cn
%}
%}}
%

\author{
    Wei~Huang,~\IEEEmembership{Member,~IEEE},
    Yong~Zeng,~\IEEEmembership{Member,~IEEE},
       and Yongming~Huang,~\IEEEmembership{Senior Member,~IEEE}
      
     %  School of Computing Science and Information, Hefei University of Technology, Hefei, China\\
       %School of Information Science and Engineering, Southeast University, Nanjing, China \\
       %E-mail: huangwei@hfut.edu.cn, huangym@seu.edu.cn
       \thanks{W. Huang is with the School of Computer Science and Information Engineering, Hefei University of Technology, Hefei 230601,  China (email: huangwei@hfut.edu.cn).
       Y. Zeng, and Y. Huang are  with the National Mobile Communications Research Laboratory, Southeast University, Nanjing 210096, China, and also with the Purple Mountain Laboratories, Nanjing 211111, China (e-mail: \{yong\_zeng, huangym\}@seu.edu.cn).}
       %\thanks{Y. Huang is the corresponding author (e-mail: huangym@seu.edu.cn).}}
% author names and IEEE memberships
% note positions of commas and nonbreaking spaces ( ~ ) LaTeX will not break
% a structure at a ~ so this keeps an author's name from being broken across
% two lines.
% use \thanks{} to gain access to the first footnote area
% a separate \thanks must be used for each paragraph as LaTeX2e's \thanks
% was not built to handle multiple paragraphs
%

% <-this % stops a space
%\thanks{M. Shell is with the Department
%of Electrical and Computer Engineering, Georgia Institute of Technology, Atlanta,}}
}
\maketitle

% As a general rule, do not put math, special symbols or citations
% in the abstract or keywords.
\begin{abstract}
This paper investigates the achievable rate region of the multiple-input single-output (MISO) interference channel aided by intelligent reflecting surfaces (IRSs). We exploit the the additional design degree of freedom provided by the coordinated IRSs to enhance the desired signal and suppress interference so as to enlarge the achievable rate region of the interference channel. To this end, we jointly optimize the active transmit beamforming at the transmitters and passive reflective beamforming at the IRSs, subject to the constant modulus constraints of reflective beamforming vectors. To address the non-convex optimization problem, we propose an iterative algorithm to optimize the transmit beamforming via second-order cone program (SOCP) and the reflective beamforming via the semi-definite relaxation (SDR). Numerical results demonstrate that the performance of  the IRS-aided  interference channel with the proposed algorithm can significantly outperform the conventional interference channel without IRS.
\end{abstract}
% Note that keywords are not normally used for peerreview papers.

% For peer review papers, you can put extra information on the cover
% page as needed:
% \ifCLASSOPTIONpeerreview
% \begin{center} \bfseries EDICS Category: 3-BBND \end{center}
% \fi
%
% For peerreview papers, this IEEEtran command inserts a page break and
% creates the second title. It will be ignored for other modes.
\IEEEpeerreviewmaketitle

\section{Introduction}
% The very first letter is a 2 line initial drop letter followed
% by the rest of the first word in caps.
%
% form to use if the first word consists of a single letter:
% \IEEEPARstart{A}{demo} file is ....
%
% form to use if you need the single drop letter followed by
% normal text (unknown if ever used by IEEE):
% \IEEEPARstart{A}{}demo file is ....
%
% Some journals put the first two words in caps:
% \IEEEPARstart{T}{his demo} file is ....
%
% Here we have the typical use of a "T" for an initial drop letter
% and "HIS" in caps to complete the first word.

 %\IEEEPARstart{I}
 Interference channel (IFC) models the communication scenario that a number of transmitters wish to send independent messages simultaneously to their respective receivers using the same channel, while causing interference to each other. It is one of the most important fundamental channel setups in wireless communication systems, especially for contemporary mobile communication networks with almost universal frequency reuse and ever-increasing node densities. The information theoretical study of IFC has a long history\cite{1055812}. The largest known achievable rate region for IFC is called as Han-Kobayashi type region \cite{1056307}, which is achieved by splitting the transmit signal of each user into common and private messages, while each receiver decodes its own designated private message and the public messages.  However, such capacity-approaching technique requires signal-level encoding/decoding cooperations among the users, which is challenging to practically implement. Alternatively, a low-complexity approach is to perform signal detection at the receivers by treating the interference as noise, while enabling transmitter-side cooperation via coordinated resource allocation strategy to enhance the achieve rate region of the IFC \cite{1237413,5895091}.\\
 \hspace*{\parindent}Pareto boundary plays an important role in characterizing the achievable rate region of  IFC, which consists of all the rate-tuples at which it is impossible to increase one user's rate without simultaneously decreasing other's.  One method for characterizing the Pareto boundary for IFC is by solving a sequence of weighted sum-rate maximization (WSRMax) problems, which are usually non-convex problems \cite{5895091}. Alternatively, by using the concept of {\it rate profile}, finding a point on the Pareto boundary of the IFC usually corresponds to the weighted-minimum-rate maximization problem, which is usually convex and hence more efficiently to solve than the  the WSRMax problems \cite{5504193}. It has been found that compared with the commonly assumed {\it proper} Gaussian signaling, i.e., the complex Gaussian signals whose in-phase and quadratic phase components are independent and identically distributed (i.i.d.), the achievable  rate region of IFC by treating interference as noise can be further improved by using {\it improper} Gaussian signaling \cite{6489066}.\\
 \hspace*{\parindent}Recently, a new type of electromagnetic surface structure called intelligent reflecting surface (IRS) or reconfigurable metasurface, is emerging as a promising component for wireless communications \cite{Smith2017Analysis,8910627}. IRS is usually composed of a large number of integrated electronic circuits that can be programmed to manipulate the incoming electromagnetic wave in a customizable manner, in which each unit of the IRS is implemented by reflective arrays that use varactor diodes with the resonant frequency electronically controlled. In wireless communication systems,  IRS can be regarded as a cost-effective implementation of passive phase shifters, which has the capability of intelligent signal reflection without any power amplifier. Thanks to its low hardware footprint, IRS can be flexibly deployed on room ceiling, buildings facades, aerial platforms \cite{LuICC2020}, even to be integrated into smart wearable devices, which brings a new design degree of freedom (DoF) to deliberately manipulate the wireless communication channels.\\
\hspace*{\parindent} Extensive research efforts on IRS-aided wireless communications have been devoted to jointly optimize the reflective beamforming  at the IRS and transmit beamforming at the multi-antenna access point \cite{9013288,8811733}. Furthermore, the IRS-aided wireless communications in various setups have been also studied, such as orthogonal frequency division multiplexing \cite{9014204}, non-orthogonal multiple access\cite{9000593} and physical layer security\cite{8743496}.\\
\hspace*{\parindent}To reap the full benefits of IRS-aided wireless communications, in this paper, we consider IRS-aided IFC, where each of the transmitter-receiver pair is aided by one IRS, as shown in Fig. 1. Intuitively, the introduction of IRS brings a new design DoF for both desired signal enhancement and interference suppression. To this end, we focus on the characterization of the Pareto boundary of the achievable rate region for the IRS-aided multiple-input single-output (MISO) IFC with interference treated as noise. Specifically, based on the concept of {\it rate-profile} \cite{5504193}, we formulate an optimization problem for Pareto boundary characterization by jointly designing active transmit beamforming at the transmitters and passive reflective beamforming at the IRSs. As the problem is non-convex, an efficient algorithm is proposed to find a high-quality suboptimal solution via block coordinate descent (BCD) method, where the transmit and reflective beamforming vectors are optimized in an alternating manner. In particular, with the fixed reflective beamforming, the optimal transmit beamforming can be obtained via second-order cone program (SOCP). On the other hand, with the fixed transmit beamforming vector, the reflective beamforming vector can be updated via the semi-definite relaxion (SDR) approach.  Numerical results show that with the proposed algorithm, the achievable rate region of IRS-aided IFC is much larger than that without IRS.\\
\begin{figure}[!t]
 \centering
 \includegraphics[width=3.2in]{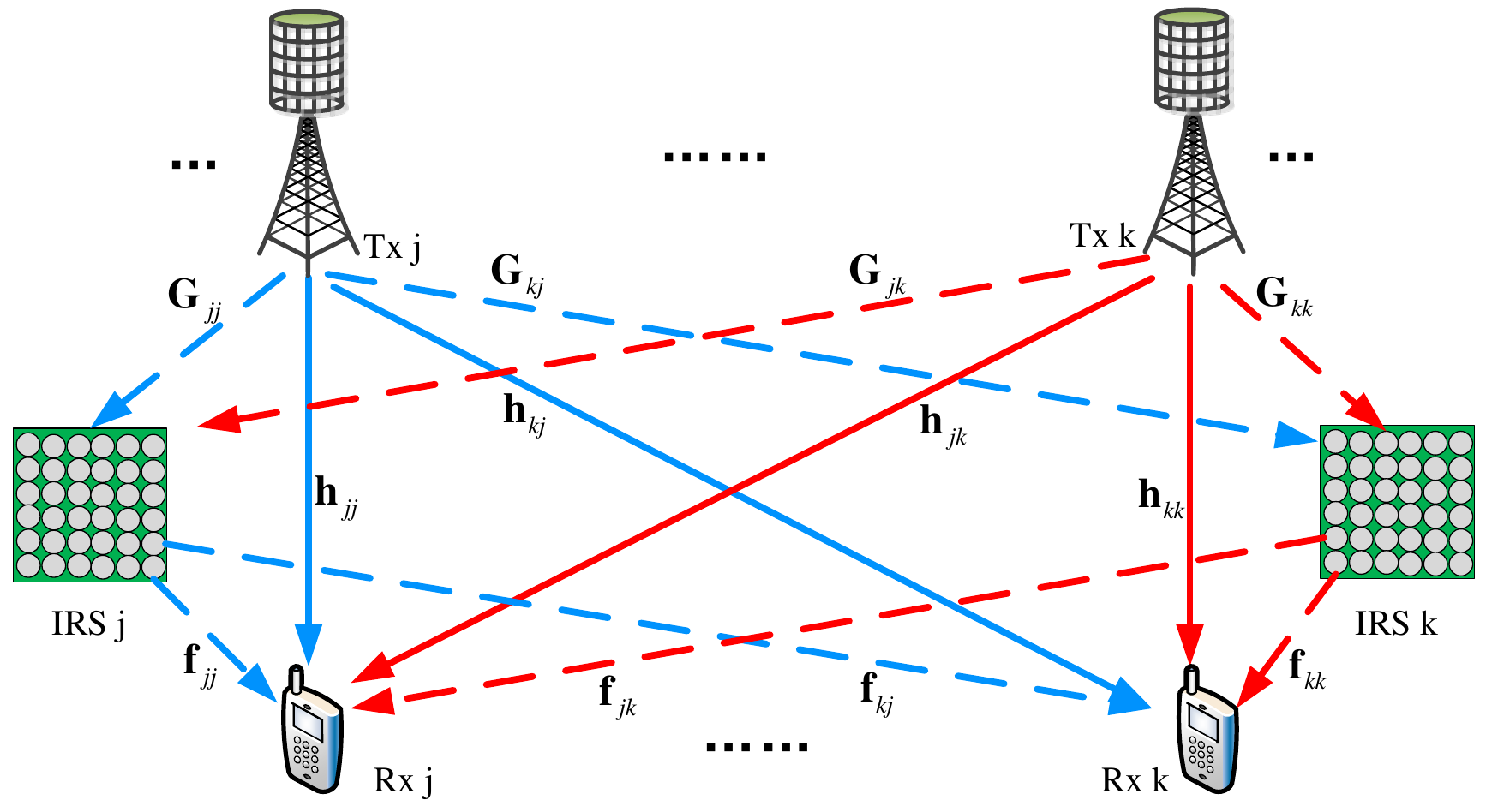} \caption{MISO interference channel aided by IRS.}     \label{fig1}
\end{figure}
\section{System Model and Problem Formulation}
As shown in Fig. 1, we consider an IRS-aided MISO IFC with $K$ transmitter-receiver pairs, each of which is aided by one IRS. Each transmitter is assumed to have $M$ antennas and each IRS has $N$ passive reflecting elements.  Denote the direct MISO channel from transmitter $j$ to receiver $k$ as ${\bf h}_{kj}\in {\mathbb C}^{M\times 1}$, where $j, k=\{1,\cdots,K\}$. Further denote the multiple-input multiple-output (MIMO) channel from transmitter $j$ to IRS $i$ as ${\bf  G}_{ij}\in {\mathbb C}^{N\times M}$, and the MISO channel from IRS $i$ to receiver $k$ as ${\bf  f}_{ki}\in {\mathbb C}^{N \times 1}$.\\
\hspace*{\parindent}Let ${\bf w}_j\in {\mathbb C}^{M\times 1}$ denote the transmit beamforming vector by transmitter $j\in \{1,....,K\}$, and $P_j$ is the maximum power of transmitter $j$, we then have $\|{\bf w}_j\|^2\leq P_j$.  We assume that each reflective element of the IRS is able to dynamically manipulate the phase of the incoming electromagnetic waves, while keeping their magnitudes unchanged. For IRS $i \in \{1,\cdots, K\}$, denote $\phi_{in} \in [0, 2\pi)$ as the phase shift introduced by the $n$th element, $n\in \{1,...,N\}$. Further, define the reflective phase shift matrix as the diagonal matrix ${\bf \Phi}_i={\rm diag}({\bf v}_i)$, where ${\bf v}_i\in {\mathbb C}^{N\times 1}=(e^{j\phi_{i1}}, \cdots,e^{j\phi_{iN}})$. Thus, the baseband complex signal received at receiver $k$ can be written as
\begin{align*}
y_k=\sum\limits_{j=1}^K{\bf h}_{kj}^{\rm H}{\bf w}_js_j+\sum\limits_{j=1}^K\sum\limits_{i=1}^K{\bf f}_{ki}^{\rm H}{\bf \Phi}_i{\bf G}_{ij}{\bf w}_js_j+z_k
\end{align*}
\begin{align}
&=\underbrace {\left({\bf h}_{kk}^{\rm H}+\sum\limits_{i=1}^K{\bf f}_{ki}^{\rm H}{\bf \Phi}_i{\bf G}_{ik}\right){\bf w}_ks_k}_{\rm {desired \mspace{3mu} signals}}\notag\\
&+\underbrace{\sum\limits_{j \neq k}\left({\bf h}_{kj}^{\rm H}+\sum\limits_{i=1}^K{\bf f}_{ki}^{\rm H}{\bf \Phi}_i{\bf G}_{ij}\right){\bf w}_js_j}_{\rm {interference}}+z_k\,,\forall k\,,
\end{align}
where $s_k$ is the information-bearing symbol for the $k$th user, $z_k$ denotes he additive white Gaussian noise (AWGN) at receiver $k$, which is assumed to be circularly symmetric complex Gaussian (CSCG) distributed with zero mean and power $\sigma^2$, i.e., $z_k\sim {\mathbb {CN}}(0,\sigma^2)$. Note that we have ignored the wireless links across different IRSs, since intuitively, the IRSs would only direct their incoming signals towards the receivers rather than other IRSs. It is observed from (1) that both the desired signal and interference of each receiver arrives not only from its own transmitter and IRS, but also from other IRSs. This provides additional DoF to enhance the achievable rate region as compared to the conventional MISO IFC without IRS. \\
\hspace*{\parindent}With the interference treated as noise at each receiver, the resulting signal-to-interference-plus-noise ratio (SINR) at receiver $k$ can be written as
\begin{align}
{\rm SINR}_k=\frac{\left|({\bf h}_{kk}^{\rm H}+\sum\limits_{i=1}^K{\bf f}_{ki}^{\rm H}{\bf \Phi}_i{\bf G}_{ik}){\bf w}_k\right|^2}{\sum\limits_{j \neq k}\left| ({\bf h}_{kj}^{\rm H}+\sum\limits_{i=1}^K{\bf f}_{ki}^{\rm H}{\bf \Phi}_i{\bf G}_{ij}){\bf w}_j\right|^2+\sigma^2}\,.
\end{align}
As ${\bf \Phi}_i$ is a diagonal matrix with diagonal elements given by vector ${\bf v}_i$, it is not difficult to see that the cascaded link from transmitter $j$ to receiver $k$ through IRS $i$ can be written as
\begin{align}
{\bf f}_{ki}^{\rm H}{\bf \Phi}_i{\bf G}_{ij}={\bf v}_i^{\rm H}{\boldsymbol \Gamma}_{kij}\,,
\end{align}
where we have defined the effective channel ${\boldsymbol \Gamma}_{kij}\in {\mathbb C}^{N\times M}={\rm diag}({\bf f}_{ki}^{\rm H}){\bf G}_{ij}$.  By substituting (3) to (2), and assuming CSCG signaling, the achievable rate of receiver $k$ is given by:
\begin{align}\label{rate1}
R_k=\log_2\left(1+\frac{\left|({\bf h}_{kk}^{\rm H}+\sum\limits_{i=1}^K{\bf v}_i^{\rm H}{\boldsymbol \Gamma}_{kik}){\bf w}_k\right|^2}{\sum\limits_{j\neq k}\left|({\bf h}_{kj}^{\rm H}+\sum\limits_{i=1}^K{\bf v}_i^{\rm H}{\boldsymbol \Gamma}_{kij}){\bf w}_j\right|^2+\sigma^2}\right)\,.
\end{align}
The achievable rate region for the IRS-aided IFC is the set of all rate-tuples $(R_1,\cdots, R_k)$ for the $K$ user pairs that can be simultaneously achieved, which can be written as
\begin{align}\label{achievable rate}
{\mathcal R}=\mathop\bigcup_{\substack{|v_{in}|=1,||{\bf w}_j||^2\le P_j\\ i,j\in\{1,...K\}, n\in \{1,...,N\}}}\{(R_1, R_2,\cdots,R_K)\}\,.
\end{align}
The outer boundary of $\cal R$ is called the Pareto boundary, which consists of all the rate-tuples at which  it is impossible to increase one user's rate without simultaneously decreasing that of other users \cite{6123786}. We are able to characterize the Pareto optimal rate-tuples based on the concept of rate profile \cite{5504193}. Specifically, any rate-tuple on the Pareto boundary of $\cal R$ can be obtained by solving the following optimization problem with a given vector ${\boldsymbol \zeta}=(\zeta_1,\cdots,\zeta_K)$:
\begin{subequations}\label{orig_formulation}
\begin{align}
\mathop{\max}_{R,\{{\bf v}_i\}_{i=1}^K,\{{\bf w}_j\}_{j=1}^K} &\quad  R\\
{\rm s.t.} \quad~~~ &  R_k\ge \zeta_kR\,, \forall k\,,\\
& |v_{in}|=1\,, \forall i,n\,,\\
& ||{\bf w}_j||^2\le P_j, \forall j\,,
\end{align}
\end{subequations}
where  $\zeta_k\geq 0$ denotes the target rate ratio between the achievable rate of receiver $k$ and the sum rate $R$, with $\sum_{k=1}^K\zeta_k=1$. Therefore, with different $\boldsymbol \zeta$, the complete Pareto boundary of the achievable rate region $\cal R$ can be characterized. Note that in the absence of the IRSs, the optimization problem (\ref{orig_formulation}) degenerates to the transmit beamforming optimization for the conventional MISO-IFC, which can be optimally solved via SOCP. However, the introduction of the IRSs renders the problem (\ref{orig_formulation})  more challenging, since the transmit and reflective beamforming vectors are coupled, which makes (\ref{orig_formulation}) non-convex and difficult to be optimally solved. In the following, we propose an efficient algorithm to solve (\ref{orig_formulation}) based on the BCD technique, for which the transmit and reflective beamforming vectors are updated alternately.
\section{Proposed Solution}
To gain some insights, we first consider the special case of (\ref{orig_formulation}) with $\zeta_k=1$ for some $k$ and $\zeta_j=0, \forall j\neq k$. This corresponds to the single-user maximum rate point for user $k$.
\subsection{Single-user Maximum Rate  Point}
 With $\zeta_j=0, \forall j\neq k$, it is obvious that we should have ${\bf w}_j=0$, $\forall j\neq k$, and $\|{\bf w}_k\|^2=P_k$. As a result, problem (\ref{orig_formulation})  reduces to
 \begin{align}\label{single formuala}
 \mathop{\max}_{\{{\bf v}_i\}_{i=1}^K,{\bf w}_k} & {\rm log}_2\left(1+\frac{1}{\sigma^2}\left|({\bf h}_{kk}^{\rm H}+\sum\limits_{i=1}^K{\bf v}_i^{\rm H}{\boldsymbol \Gamma}_{kik}){\bf w}_k\right|^2\right)\notag\\
{\rm s.t.} \quad &  |v_{in}|=1\,, \forall i,n\,,\notag\\
& ||{\bf w}_k||^2= P_k.
 \end{align}
It is not difficult to see that for any given reflective beamforming vectors $\{{\bf v}_i\}_{i=1}^K$, the optimal transmit beamforming vector ${\bf w}_k^\star$ is given by  the maximum ratio transmit (MRT) beamforming with the effective channel formed by a supposition of the direct channel and the reflective channels, which is given by
 \begin{align}\label{opti trans}
  {\bf w}_k^\star=\sqrt{P_k}\frac{{\bf h}_{kk}+\sum_{i=1}^K{\boldsymbol \Gamma}_{kik}^{\rm H}{\bf v}_i}{\left \|{\bf h}_{kk}+\sum_{i=1}^K{\boldsymbol \Gamma}_{kik}^{\rm H}{\bf v}_i\right\|}\,.
 \end{align}
With (\ref{opti trans}), the resulting SNR for user $k$ is only a function of the reflective beamforming vectors $\{{\bf v}_i\}_{i=1}^K$, which is
\begin{align}\label{snr}
{\rm SNR}_k=\frac{P_k}{\sigma^2}\left\|{\bf h}_{kk}+\sum_{i=1}^K{\boldsymbol \Gamma}_{kik}^{\rm H}{\bf v}_i \right\|^2=\frac{P_k}{\sigma^2}f(\{{\bf v}_i\})\,.
\end{align}
\hspace*{\parindent}As a result, problem (\ref{single formuala})  reduces to the SNR maximization problem via reflective beamforming optimization:
\begin{align}\label{formu_snr}
\mathop{\max}_{\{{\bf v}_i\}_{i=1}^K} \mspace{3mu}{\rm SNR}_k, \quad {\rm s.t.} \mspace{3mu}|v_{in}|=1\,, \forall i,n\,.
\end{align}
\hspace*{\parindent}The non-convex unit-magnitude constraints on the elements of the reflective beamforming vectors in (\ref{formu_snr}) make it difficult to find  the optimal solution efficiently. In the following, we propose a heuristic algorithm to find an efficient local optimal solution. Specifically, we aim to extract the contribution from each individual element $v_{in}$ to the ${\rm SNR}_k$ in (\ref{snr}), with all other elements fixed, and update each element alternately via the classic coordinate ascent method\cite{7389996}. To proceed, let ${\boldsymbol \Gamma}_{kik}^{\rm H}=[{\bf r}_1,{\bf r}_2,\cdots,{\bf r}_{N}]$, where ${\bf r}_n\in {\mathbb C}^{M\times 1}$ is the $n$th column of matrix ${\boldsymbol \Gamma}_{kik}^{\rm H}$. Then, $f(\{{\bf v}_i\})$ in (\ref{snr}) can be expressed as
\begin{align}\label{maximizer}
f(\{{\bf v}_i\})&=\|\underbrace{{\bf h}_{kk}+\sum_{i'\neq i}{\boldsymbol \Gamma}_{ki'k}^{\rm H}{\bf v}_{i'}}_{\triangleq {\bf g}_i}+ {\boldsymbol \Gamma}_{kik}^{\rm H}{\bf v}_i\|^2\notag\\
&=\|\underbrace{{\bf g}_i+\sum_{n'\neq n}{\bf r}_{n'}e^{j\phi_{in'}}}_{\triangleq {\bf \bar g}_{in}}+{\bf r}_ne^{j\phi_{in}}\|^2=||{\bf \bar g}_{in}+{\bf r}_ne^{j\phi_{in}}||^2\notag\\
&=||{\bf \bar g}_{in}||^2+||{\bf r}_n||^2+2{\rm Re}\{e^{j\phi_{in}}{\bf \bar g}_{in}^{\rm H}{\bf r}_n\}\,.
\end{align}
With all elements of $\{{\bf v}_i\}_{i=1}^K$ fixed except  $e^{j\phi_{in}}$, the optimal value for $\phi_{in}$ should maximize ${\rm Re}\{e^{j\phi_{in}}{\bf \bar g}_i^{\rm H}{\bf r}_n\}$, which is $\phi_{in}^\star=-\angle{\bf \bar g}_{in}^{\rm H}{\bf r}_n$. Now, we are able to devise an iterative algorithm to and alternately update each entry $v_{in}=e^{j\phi_{in}^\star}$ with all other elements fixed. Since the ${\rm SNR}_k$ in (\ref{snr}) increases (at least non-decreasing) at each iteration, the algorithm is guaranteed to converge. \\
\hspace*{\parindent}It is noted that for the $K$-user MISO-IFC aided by IRS, for the special case with single-user maximum rate point, all the transmitters except transmitter $k$ will keep silence, and all IRSs work cooperatively to enhance the effective MISO channel of user $k$. Therefore, compared with the conventional IFC, the single-user maximum rate point is guaranteed to be improved in the presence of IRS.
\subsection{Iterative Transmit and Reflective Beamforming Optimization Design}
In this subsection, we consider the general optimization problem (\ref{orig_formulation}) for multi-user MISO-IFC aided by IRS. An iterative optimization algorithm based on the BCD technique is proposed, where the transmit active beamforming and reflec- tive passive beamforming are updated alternately with the other fixed. First, consider the transmit beamforming optimization problem  with given reflective beamforming vectors $\{{\bf v}_i\}_{i=1}^K$. In this case,  the effective MISO channel from transmitter $j$ to reciever $k$, denoted as ${\bf g}_{kj}\in {\mathbb C}^{M\times 1}$, can be written
\begin{align*}
{\bf g}_{kj}=\left({\bf h}_{kj}+\sum_{i=1}^K{\boldsymbol \Gamma}_{kij}^{\rm H}{\bf v}_i\right)\,.
\end{align*}
As a result, the sub-problem for transmit beamforming optimization of problem (\ref{orig_formulation}) reduces to
\begin {subequations}\label{relax power}
\begin{align}
\mathop{\max}_{R,\{{\bf w}_j\}_{j=1}^K} &\quad  R\\
{\rm s.t.} \quad  &\log_2\left(1+\frac{\left|{\bf g}_{kk}^{\rm H}{\bf w}_k\right|^2}{\sum_{j\neq k}\left|{\bf g}_{kj}^{\rm H}{\bf w}_j\right|^2+\sigma^2}\right)\ge \zeta_kR\,,\forall k\,,\\
&  ||{\bf w}_j||^2\le P_j, \forall j\,.
\end{align}
\end{subequations}
Problem (\ref{relax power}) reduces to the beamforming optimization problem of the conventional MISO IFC, which can be optimally solved via SOCP together with bisection search method [8]. Specifically, for any given rate target $R > 0$, we have the following feasibility problem:
\begin {subequations}\label{last power}
\begin{align}
{\rm Find:}\quad & \{{\bf w}_j\}_{j=1}^K,\\
{\rm s.t.}\quad  &\frac{|{\bf g}_{kk}^{\rm H}{\bf w}_k|^2}{ (2^{\zeta_kR}-1)}\ge \sum\limits_{j\neq k}\left|{\bf g}_{kj}^{\rm H}{\bf w}_j\right|^2+\sigma^2\,,\forall k\,,\\
& ||{\bf w}_j||^2\le P_j, \forall j\,.
\end{align}
\end{subequations}
If $R$ is feasible to problem (\ref{last power}), then, the optimal value of problem (\ref{relax power}), denoted as $R^\star$, is no smaller than the given value $R$, i.e., $R^\star \geq R$; otherwise, we have $R^{\star} < R$. Hence, the optimal solution $\{R^\star, \{{\bf w}_j^\star\}_{j=1}^K\}$ to problem (\ref{relax power}) can be obtained via bisection search over $R$ by solving a sequence of the feasibility problem  (\ref{last power}). Moreover, it is noted that without loss of optimality, a common phase shift can be applied to ${\bf w}_k$ so that ${\bf g}_{kk}^{\rm H}{\bf w}_k$ is a real value for all $k$. Therefore, constraints (\ref{last power}b) can be recast as second order cone (SOC) constraints as
\begin{align}\label{soc constraint}
\frac{{{\mathop{\rm Re}\nolimits} \left( {{\bf{g}}_{kk}^{\rm{H}}{{\bf{w}}_k}} \right)}}{{{2^{{\zeta _k}R}} - 1}} \ge \left\| \begin{array}{l}
{\bf{g}}_{k1}^{\rm{H}}{{\bf{w}}_1}\\
\quad \vdots \\
{\bf{g}}_{k(k - 1)}^{\rm{H}}{{\bf{w}}_{k - 1}}\\
{\bf{g}}_{k(k + 1)}^{\rm{H}}{{\bf{w}}_{k + 1}}\\
 \quad\vdots \\
{\bf{g}}_{kK}^{\rm{H}}{{\bf{w}}_K}\\
{\sigma}
\end{array} \right\|\,,\forall k\,,
\end{align}
 which are convex constraints. Therefore, problem (\ref{last power}) can be further written as:
 \begin{align}\label{last socp}
 {\rm Find:}\quad  \{{\bf w}_j\}_{j=1}^K\,,\quad
{\rm s.t.}\quad   (\ref{last power}c) \mspace{6mu} {\rm and} \mspace{6mu}  (\ref{soc constraint})
 \end{align}
Problem (\ref{last socp}) is an SOCP problem, which can be efficiently solved by standard convex optimization solvers\cite{w1995x}. Thus, the solution of optimal transmit beamforming to problem (\ref{relax power}) is obtained via solving the SOCP problem together with a bisection search over $R$.\\
\hspace*{\parindent}Next, we focus on optimizing the reflective beamforming vectors $\{{\bf v}_i\}_{i=1}^K$ with fixed transmit beamforming. With any given $\{{\bf w}_j\}_{j=1}^K$, constraints (\ref{orig_formulation}b) can be rewritten as
\begin{align}\label{formula constraint}
\log_2\left(1+\frac{\left|a_{kk}+\sum\limits_{i=1}^K{\bf v}_i^{\rm H}{\boldsymbol \gamma}_{kik}\right|^2}{\sum_{j\neq k}\left| a_{kj}+\sum\limits_{i=1}^K{\bf v}_i^{\rm H}{\boldsymbol \gamma}_{kij}\right|^2+\sigma^2}\right)\ge \zeta_kR\,,
\end{align}
where $ a_{kj}\triangleq{\bf h}_{kj}^{\rm H}{\bf w}_j$ and ${\boldsymbol \gamma_{kij}}\triangleq{\boldsymbol \Gamma}_{kij}{\bf w}_j\in {\mathbb C}^{N\times 1}$.  Define
\begin{align*}
{\bf \bar v}\buildrel \Delta \over = \left[ \begin{array}{l}
{1}\\
{{\bf v}_1}\\
\vdots \\
{{\bf v}_K}
 \end{array} \right]\in {\mathbb C}^{(KN+1)\times 1}\,,
{\bf b}_{kj}\buildrel \Delta \over = \left[ \begin{array}{l}
a_{kj}\\
{{\boldsymbol \gamma}_{k1j}}\\
\vdots \\
{{\boldsymbol \gamma}_{kKj}}
 \end{array} \right]\in {\mathbb C}^{(KN+1)\times 1}\,.
\end{align*}
Then, the sub-problem for reflective beamforming optimization of problem (\ref{orig_formulation}) can be written as
\begin{subequations} \label{same power}
\begin{align}
\mathop{\max}_{{\bf \bar v},R}\quad &R\\
{\rm s.t.}\quad&\log_2\left(1+\frac{|{\bf b}_{kk}^{\rm H}{\bf \bar v}|^2}{\sum_{j\neq k}|{\bf b}_{kj}^{\rm H}{\bf \bar v}|^2+\sigma^2}\right)\ge \zeta_kR\,, \forall k\,,\\
&{\bar v}_1=1\,,\\
&|{\bar v}_{n}|=1\,,n=2,3,\cdots,KN+1\,.
\end{align}
\end{subequations}
Problem (\ref{same power}) is non-convex, due to the non-concave rate expression with respect to the reflective beamforming vectors and the non-convex unit-magnitude constraints. To address the problem, we leverage the celebrated SDR method to find a high-quality approximate solution\cite{4443878}. To proceed, define a rank-1 positive semi-definite matrix  ${\bf V}={\bf \bar v}{\bf \bar v}^{\rm H}$. Then, problem (\ref{same power}) is further reformulated as 
\begin{subequations}\label{relax0}
\begin{align}
\mathop{{\max}}_{R,{\bf V}}\quad &R\\
{\rm s.t.}\quad &\frac{{\rm Tr}({\bf Q}_{kk}{\bf V})}{\sum_{j\neq k}{\rm Tr}({\bf Q}_{kj}{\bf V})+\sigma^2}\ge 2^{\zeta_kR}-1\,,\\
&|{\bf V}_{nn}|=1\,, n={1,2\cdots, KN+1}\,,\forall k\,,\\
& {\rm rank} ({\bf V})= 1\,,\\
& {\bf V}\succeq {\bf 0}\,,
\end{align}
\end{subequations}
where ${\bf Q}_{kj}={\bf b}_{kj}{\bf b}_{kj}^{\rm H}$. For the given rate target $R>0$, we have the following feasibility problem:
\begin{subequations}\label{relax1}
\begin{align}
{\rm Find}:\quad &{\bf V}\\
{\rm s.t.}\quad &\frac{{\rm Tr}({\bf Q}_{kk}{\bf V})}{ (2^{\zeta_kR}-1)}\ge\sum\limits_{j\neq k}{\rm Tr}({\bf  Q}_{kj}{\bf V})+\sigma^2\,, \forall j,k\,,\\
&(\ref{relax0}c), \mspace{3mu}(\ref{relax0}d) \mspace{3mu}{\rm and}\mspace{3mu} (\ref{relax0}e) \,.
\end{align}
\end{subequations}
Note that (\ref{relax1}) is still non-convex, due to the rank-one constraints. By dropping the rank-one constraints, we have the following relaxed problem:
\begin{align}\label{relax2}
{\rm Find}:\quad {\bf V}\,,\quad
{\rm s.t.}\quad(\ref{relax0}c), \mspace{3mu} (\ref{relax0}e) \mspace{3mu}{\rm and}\mspace{3mu}(\ref{relax1}b)\,.
\end{align}
\hspace*{\parindent}Problem (\ref{relax2}) is a convex SDP problem, which can be solved directly. However, due to the relaxation, the resulting $\bf V$ from (\ref{relax2}) may not be rank-one matrix. To this end, we apply the standard Gaussian randomization steps\cite{4443878} to construct a rank-one solution to problem (\ref{relax1}). Therefore, problem (\ref{relax0}) is solved with bisection search over $R$, by solving a sequence of the SDP problem together with Gaussian randomization to construct rank-1 matrix. \\
\hspace*{\parindent}In summary, the proposed algorithm to solve problem (6)  is presented in Alg.~1. In Alg.~1, for the different weight vector $\boldsymbol \zeta$, the transmit and reflective beamforming vectors are optimized alternately. By varying $\boldsymbol \zeta$, the resulting rate tuples constitute the Pareto boundary of the achievable rate region for the multi-user MISO IFC aided by the coordinated IRS.
%\begin{figure}[!t]
%\includegraphics[width=3.5in]{achieve_snr.pdf}
 %\caption{Max-min rate versus SNR for three-user MISO IFC.}
 %\label{fig3}
%\end{figure}
\section{Numerical Results}
In this section, we provide an example to evaluate the performance of the proposed algorithms. We consider a two-user IFC with $K=2$ and $M=32$, and the two transmitters are located at $(0,50m)$ and $(50m, 50m)$, respectively, and the receivers are located at  $(0,0)$ and $(50m,0)$, respectively. The two IRSs with $N=256$ elements are placed at two random locations between the transmitters and receivers. We assume the distance-dependent path loss model, i.e.,  $Q_{L}=C_0\left({d}/{d_0}\right)^{\beta}$, where $C_0$ denotes the path loss at the reference distance of $d_0=1$ meter, $\beta$ is the path loss exponent and $d$ denotes the  individual link distance. For the transmitter-receiver, transmitter-IRS and IRS-receiver links, set to $3.6$, $2$ and $2.5$, respectively. Moreover, for the SDR approach, $3000$ Gaussion randomizations are used to construct rank-1 matrix for any obtained higher-rank matrices from (\ref{relax2}). The following benchmark schemes are also considered: \\
$\quad~\bullet$ \textbf{Scheme 1} (\emph{Random reflective beamforming}): The phase shift of each IRS element $v_{in}$ is set as a random value uniformly distributed in $[0,2\pi)$.\\
$\quad~\bullet$ \textbf{Scheme 2} (\emph{Without IRS}): This corresponds to the Pareto optimal  solution of the conventional MISO-IFC without IRS.\\
\hspace*{\parindent}In Fig.~2, the achievable rate regions for an example two-user MISO IFC for various schemes are plotted with the ${\rm SNR}=20 \mspace{2mu}{\rm dB}$. The figure reveals that the achievable rate region for IRS-aided MISO IFC with the proposed design is much larger than that for the conventional MISO IFC. Moreover, it is observed that the IRS-aided MISO IFC with random beamforming may perform even worse  than that of the IFC without IRS. This is because the reflective links may even weaken the signal strength of direct link, if the reflective phase shift is not properly designed.
\begin{algorithm}
\caption{Iterative transmit and reflective beamforming optimization for problem (6)}
\label{alg1}
\begin{algorithmic}[1]
\STATE Initialize:  threshold $\epsilon>0$, $\{{\bf v}_i\}_{i=1}^K$, $R_{\rm min}=0$ and $R_{\rm max}$  \\
  to a sufficiently large value. Let $R_L=R_{\rm min}$ ;
\STATE \textbf{Repeat}
\STATE \quad Let $R_U=R_{\rm max}$;\\
\STATE \quad \textbf{Repeat}
\STATE \quad~ Let $R=(R_L+R_U)/2$;\\
\STATE \quad~ With the given $\{{\bf v}_i\}_{i=1}^K$, solve the SOCP problem (\ref{last socp})  \\
\quad~  and denote the optimal solution as  $\{{\bf w}_j^\star\}_{j=1}^k$;\\
\STATE \quad~ \textbf{if} (\ref{last socp}) is feasible, set $R_L=R$, ${\bf w}_j={\bf w}_j^\star,\forall j$
\STATE \quad~ \textbf{else} set $R_U=R$;
\STATE \quad \textbf{Until}  $R_U-R_L\le \epsilon$;
\STATE \quad Reset $R_U=R_{\rm max}$;\\
\STATE \quad \textbf{Repeat}
\STATE \quad~ $R=(R_L+R_U)/2$;\\
\STATE \quad With the fixed $\{{\bf w}_j\}_{j=1}^K$, solve problem (\ref{relax2}) and denote\\
 \quad   the optimal solution as ${\bf V}^\star$;
\STATE \quad~ \textbf{if} (\ref{relax2})  is feasible
\STATE \quad~~\textbf{if} ${\rm rank}({\bf V}^\star)=1$ 
\STATE \quad~~~we have ${\bf V}^\star={\bf \bar v}{\bf \bar v}^{\rm H}$, $\{{\bf v}_i\}_{i=1}^K$ and resulting rate $R$,
\STATE \quad~~\textbf{else}
\STATE \quad~~~Performing Gaussian randomization to obtain a rank-1\\ 
\quad~~~vector $\bf \bar v$, $\{{\bf v}_i\}_{i=1}^K$, and resulting rate $R$. 
\STATE \quad~~Set $R_L=R$,\\
\STATE \quad~ \textbf{else} set $R_U=R$;
\STATE \quad \textbf{Until}  $R_U-R_L\le \epsilon$;
\STATE  \textbf{Until} the increase of objective function is smaller than $\epsilon$
\STATE \textbf{Output:} Solution $\{{\bf w}_j\}_{j=1}^K$, $\{{\bf v}_i\}_{i=1}^K$
\end{algorithmic}
\end{algorithm}
\begin{figure}[!t]
 \includegraphics [width=2.7in]{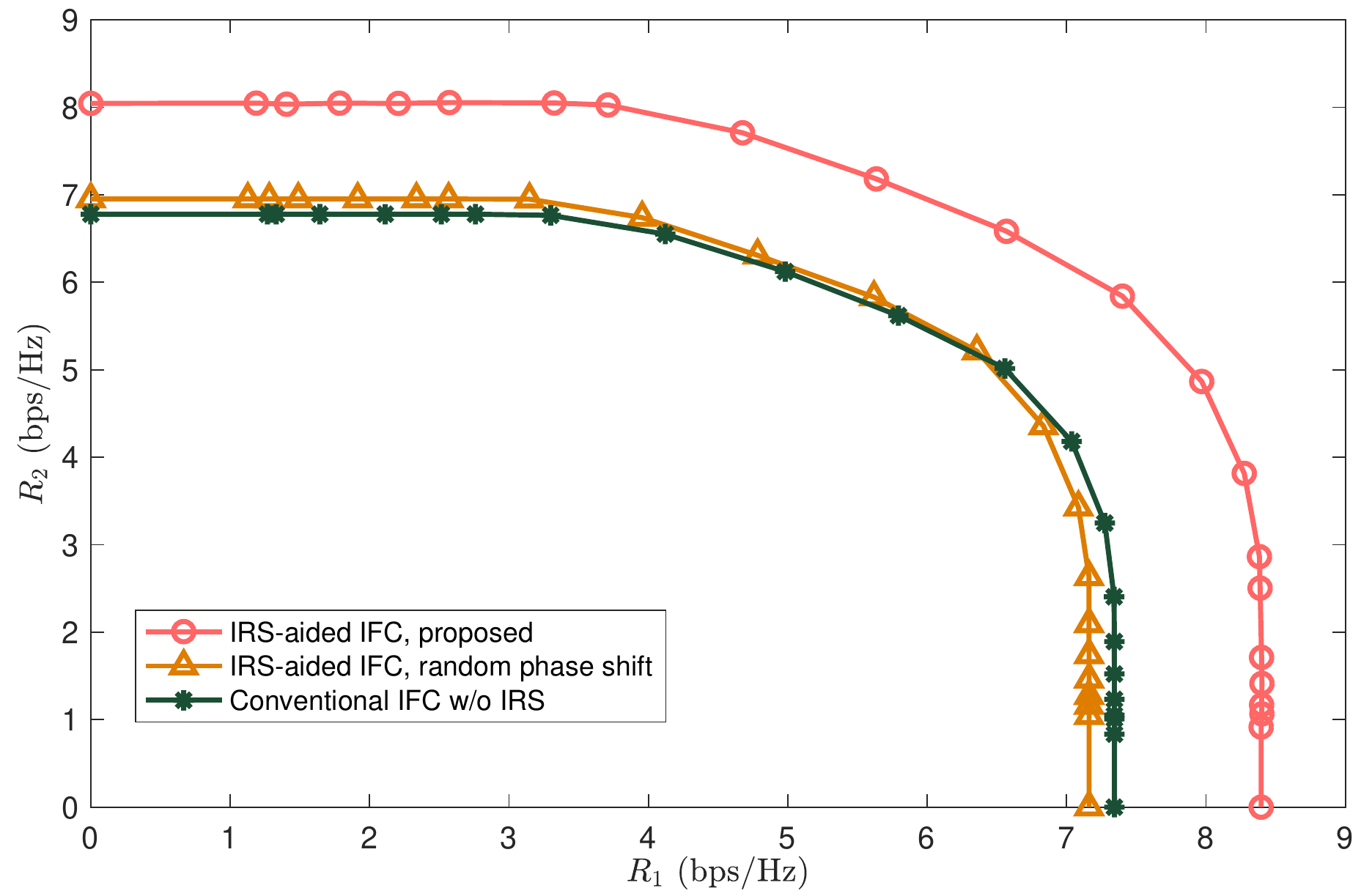}
 \caption{Achievable rate region of  two-user MISO IFC with or without IRS.}
 \label{fig2}
\end{figure}
\section{Conclusion}
In this paper, we studied the achievable rate region of the multi-user MISO IFC aided by coordinated IRSs. By leveraging the additional design DoF provided by  multiple IRSs, we proposed an iterative transmit and reflective beamforming design scheme to characterize the achievable rate region of IFC, based on SOCP and SDR optimization techniques. Numerical results were provided to demonstrate that the achievable rate region of IRS-aided IFC outperforms that of the conventional IFC, if the beamforming is properly designed.
% if have a single appendix:
%\appendix[Proof of the Zonklar Equations]
% or
%\appendix  % for no appendix heading
% do not use \section anymore after \appendix, only \section*
% is possibly needed
% use appendices with more than one appendix
% then use \section to start each appendix
% you must declare a \section before using any
% \subsection or using \label (\appendices by itself
% starts a section numbered zero.)
%
%\appendices
%Appendix one text goes here.
% you can choose not to have a title for an appendix\bigcap
% if you want by leaving the argument blank
%\section{Appendix}
%Appendix two text goes here.
% use section* for acknowledgement
%\section*{Acknowledgment}
%The authors would like to thank...
% Can use something like this to put references on a page
% by themselves when using endfloat and the captionsoff option.
\ifCLASSOPTIONcaptionsoff
  \newpage
\fi
\bibliographystyle{IEEEtran}
% argument is your BibTeX string definitions and bibliography database(s)
\bibliography{irs_bio}
\end{document}